**Signatures of a light-induced exciton condensate exhibiting BEC-BCS crossover**

Khanh Duy Nguyen[1], Gabriele Berruto[1], Yunhe Bai[1], Thomas Marchese[1], Woojoo Lee[1,2], Haoran Lin[1], Jiangang Yang[1], Chong Liu[1], Y. Shirley Meng[1], Shuolong Yang[1,*]

[1]*Pritzker School of Molecular Engineering, The University of Chicago, Chicago, IL 60637, USA*

[2]*Department of Physics, Gachon University, Seongnam 13120, Republic of Korea*

*Corresponding author. Email: yangsl@uchicago.edu

## Abstract

Exciton condensates provide a platform to study quasiparticle pairing, Bose-Einstein condensation–Bardeen-Cooper-Schrieffer (BEC-BCS) crossover, and excitonic topological phenomena. Achieving a nonequilibrium exciton condensate allows the ultimate tunability of these emergent phenomena. Yet, evidence of a light-induced, nonequilibrium exciton condensate and its BEC-BCS crossover remains elusive. Here, we use time- and angle-resolved photoemission spectroscopy to demonstrate signatures of a non-equilibrium exciton condensate and its BEC-BCS crossover in monolayer $MnBi_2Te_4$. Following optical excitation, a distinctive hole-like dispersion representing excitons emerges and persists for > 20 ps. Strikingly, energy-domain sharpening in the valence band occurs 2 ps after time zero and exhibits a sharp onset at a threshold pump fluence of 0.84 mJ/cm$^2$. The delayed and strongly nonlinear response is difficult to reconcile with transient field effects or conventional carrier-induced band shifts but is consistent with a model of exciton condensation governed by a Berezinskii-Kosterlitz-Thouless transition. The estimated threshold exciton density agrees quantitatively with the Nelson-Kosterlitz critical density. At higher fluences, the exciton feature develops a camel-back-shaped dispersion, consistent with the BEC-BCS crossover in the condensate framework. Our work establishes ultrathin $MnBi_2Te_4$ as a model system for studying nonequilibrium exciton condensates with a connection to superconductivity and exciton-driven topological phases.

## Introduction

Along with other charge-neutral electronic excitations in many-body physics, excitons – electron-hole pairs obeying bosonic statistics – have attracted considerable attention due to their versatile roles in coupling to electronic orders and to external or emergent gauge fields[1]. A macroscopic occupation of the excitonic ground state leads to an exciton condensate, realized in delicately constructed 2D material stacks[2,3]. By separating electron and hole layers, these heterostructures enable fine control over the exciton density and coupling strength, realizing dissipationless transport and BEC-BCS crossover[3]. Excitonic bandgaps in equilibrium have been reported in materials such as monolayer $WTe_2$[4,5] and bulk $Ta_2Pd_3Te_5$[6,7], but such symmetry-breaking phases may be interfered by lattice distortions[8].

At the same time, optically driven nonequilibrium excitonic systems have not been explored in-depth to realize exciton condensates or the BEC-BCS crossover. Notably, the BEC-BCS crossover describes the continuous, unitary evolution of a quantum fluid from a low-density regime of tightly bound, localized boson pairs to a high-density regime of weakly bound, extensively overlapped fermion pairs[9–11]. Realizing *in situ*, dynamical, and deterministic control of the BEC-BCS crossover within a solid-state platform remains an outstanding challenge. Nonequilibrium exciton condensates, if realized, offer an ideal venue to test the BEC-BCS physics: optical driving can continuously tune the exciton density and thus the crossover from localized bosonic physics to collective fermionic pairing, paving the way for dissipationless optoelectronics. Although nonequilibrium exciton formation in monolayer[12] and hetero-bilayer[13] transition metal dichalcogenides has been observed using time- and angle-resolved photoemission spectroscopy (trARPES), no conclusive evidence in these measurements has been shown for exciton condensates or BEC-BCS crossover despite a number of theoretical predictions[14–16] and intense debate[17,18].

Here we report spectroscopic signatures suggesting a nonequilibrium exciton condensate in one-septuple-layer (1-SL) $MnBi_2Te_4$ (MBT) and the BEC-BCS crossover via continuous tuning of the exciton density. Multilayer MBT is a magnetic topological insulator with A-type antiferromagnetism and strong spin-orbit coupling, hosting the quantum anomalous Hall (QAH) effect[19], axion insulating states[20], and nontrivial quantum geometry[21]. Monolayer MBT is a magnetic semiconductor with almost degenerate direct and indirect bandgaps near 0.5 eV[22,23]. It is expected to host strongly bound excitons as dielectric screening is reduced in the 2D limit. We grow 1-SL MBT using molecular beam epitaxy (MBE) and employ *in situ* trARPES with 0.5 eV pumping to observe exciton formation. Following optical excitation, we observe excitons persisting for over 20 ps within the material's gapped electronic structure. By systematically tracking the ARPES spectrum as a function of pump fluence, we observe two rapid, nonlinear changes near a common threshold fluence, suggestive of collective excitonic behaviors. First, across a critical pump fluence of ~0.84 mJ/cm$^2$, the valence band (VB) sharpens after 2 ps, leading to a peak in the differential spectrum. This VB peak stays robust for at least 10 ps even when the exciton density is substantially depleted. The delayed emergence of the VB peak, its distinct lifetime, and the abrupt onset at the threshold fluence are inconsistent with surface photovoltage (SPV)[24], optical Floquet[12] or Stark shifting[25], exciton-Floquet effect[26], or carrier- or exciton-driven bandgap renormalization[27,28]. Based on a model of exciton condensation[15] incorporating phase fluctuations[29], we attribute the VB peak to a spectroscopic feature of the condensed phase, analogous to the coherent peaks observed in unconventional superconductors[9,30–32]. In this picture, the onset behavior follows the Nelson-Kosterlitz (NK) jump of superfluid density in a Berezinskii-Kosterlitz-Thouless (BKT) transition[33,34], where rapid growth of the phase-correlation length near the transition leads to an emerged VB peak. Our observed critical density quantitatively matches the NK criterion. Second, above the critical fluence, the excitonic dispersion abruptly changes from hole-like to a camel-back shape. In the condensate framework, this dispersion change suggests a BEC-BCS crossover. As monolayer MBT constitutes the fundamental building block of a vast family of magnetic topological phases, the tunable, non-equilibrium exciton condensate and its BEC-BCS crossover suggested by our work establish an experimental bridge between excitonic correlation, superconductivity, and topological physics, opening opportunities for optical control of correlated topological quantum matter.

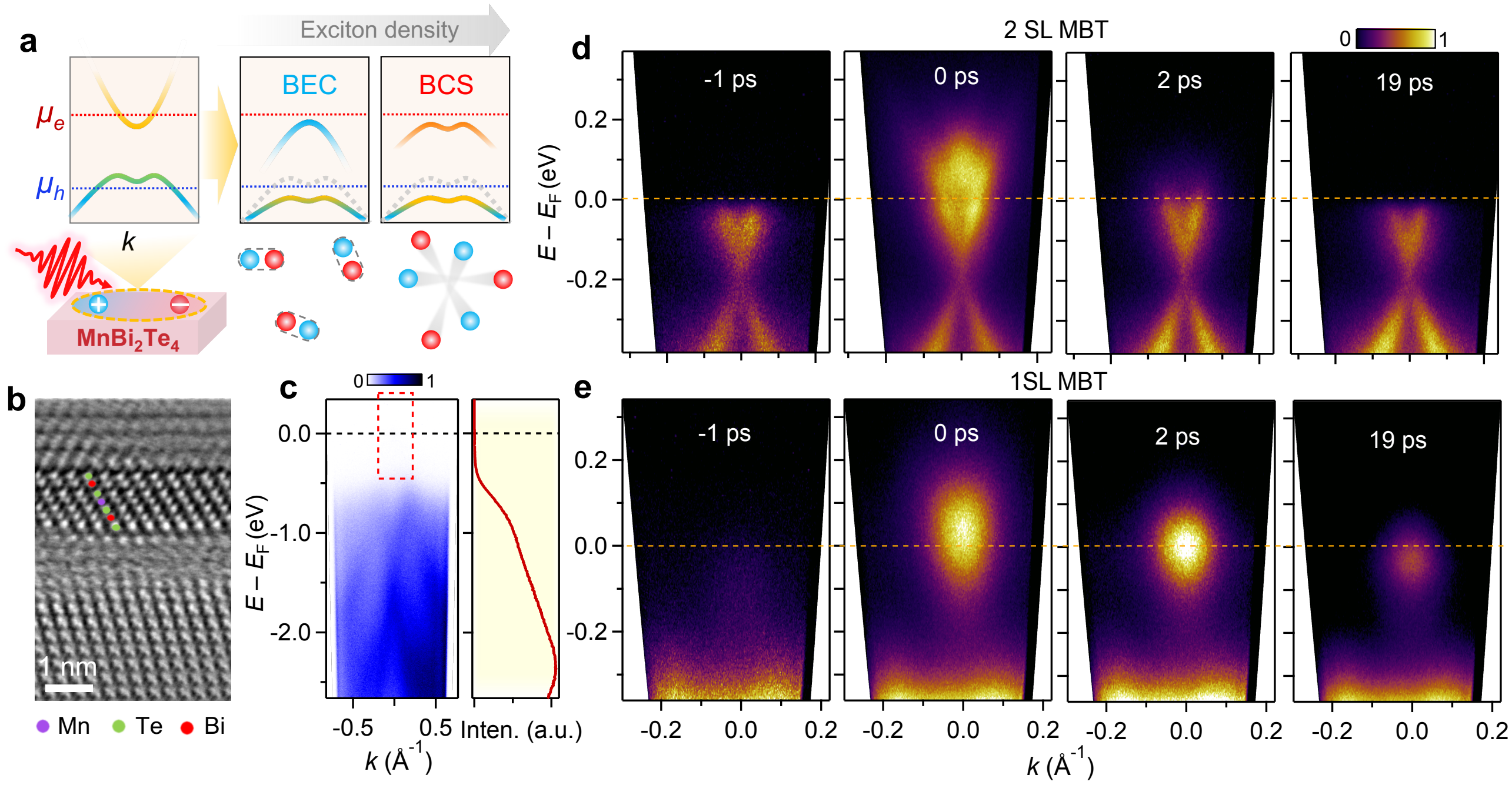

**Fig. 1| MBE-grown $MnBi_2Te_4$ (MBT) as a platform to study excitonic states**. **a,** Schematic of the trARPES experiment on MBT using a 0.5 eV pump. Band structures represent the initial excited electronic states which transform into the Bose-Einstein condensate (BEC) regime or the Bardeen-Cooper-Schrieffer (BCS) regime of exciton condensates. **b,** Scanning transmission electron microscopy image of a 1-SL MBT film grown on top of a Si (111) substrate and capped by Te, viewed along the $[1\bar{1}0]$ direction of Si. **c,** Static band structure along Γ–K measured by 21.2 eV photons showing valence band (VB) dispersions (left) and an indirect band gap around 0.5 eV evidenced by the integrated density of states (right). The red dashed box represents the energy-momentum range of the trARPES spectra in **d-e**. **d,** TrARPES spectra at representative delays for a 2-SL MBT/STO sample, featuring excited topological surface states (TSSs) as normally observed in topological insulators. **e**, TrARPES spectra for a 1-SL MBT/Si sample. While TSSs are absent, a round-shaped excitonic feature appears and persists to at least 19 ps. The cryostat base temperature is 10 K.

## Results

MBT was grown layer-by-layer on $SrTiO_3$ (STO) (111) and Si (111) substrates. Schematically, the formation of exciton condensates is expected to yield distinct in-gap features dependent on the BEC or BCS regime[10,35,36] (Fig. 1**a**). The atomically precise synthesis is confirmed by scanning transmission electron microscopy (Fig. 1**b**). The VB dispersion of the 1-SL sample generally agrees with previous reports on 1-SL MBT films, featuring an indirect band gap[22,23]. The VB top is observed at ~0.4 eV below the Fermi level ($E_F$), as shown in Fig. 1**c**. The band structures for 2-SL MBT/STO and 1-SL MBT/Si are shown in Fig. 1**d** and **e**. The topological surface states (TSSs) appear in thicker samples as a Dirac cone at Γ (Fig. 1**d**). We perform trARPES experiments with 0.5 eV pumping and 6 eV probing on 1- and 2-SL samples to distinguish their spectroscopic responses to excitations. 2-SL MBT exhibits conventional population dynamics of the TSSs[37–39], in which the unoccupied bands are instantly filled by pumping, then cooled down within a few ps to a near-equilibrium state (Fig. 1**d**). In contrast, 1-SL MBT exhibits a round-shaped nonequilibrium feature at Γ, which is localized in the energy-momentum space and lasts at least 20 ps after the pump arrival (Fig. 1**e**). We attribute this feature to excitons, whose detailed dispersion and time evolution will be discussed in subsequent parts of this article. The excitonic feature is also observed in 1-SL MBT/STO, although the exact binding energy is substrate-specific (Extended Data Fig. 1).

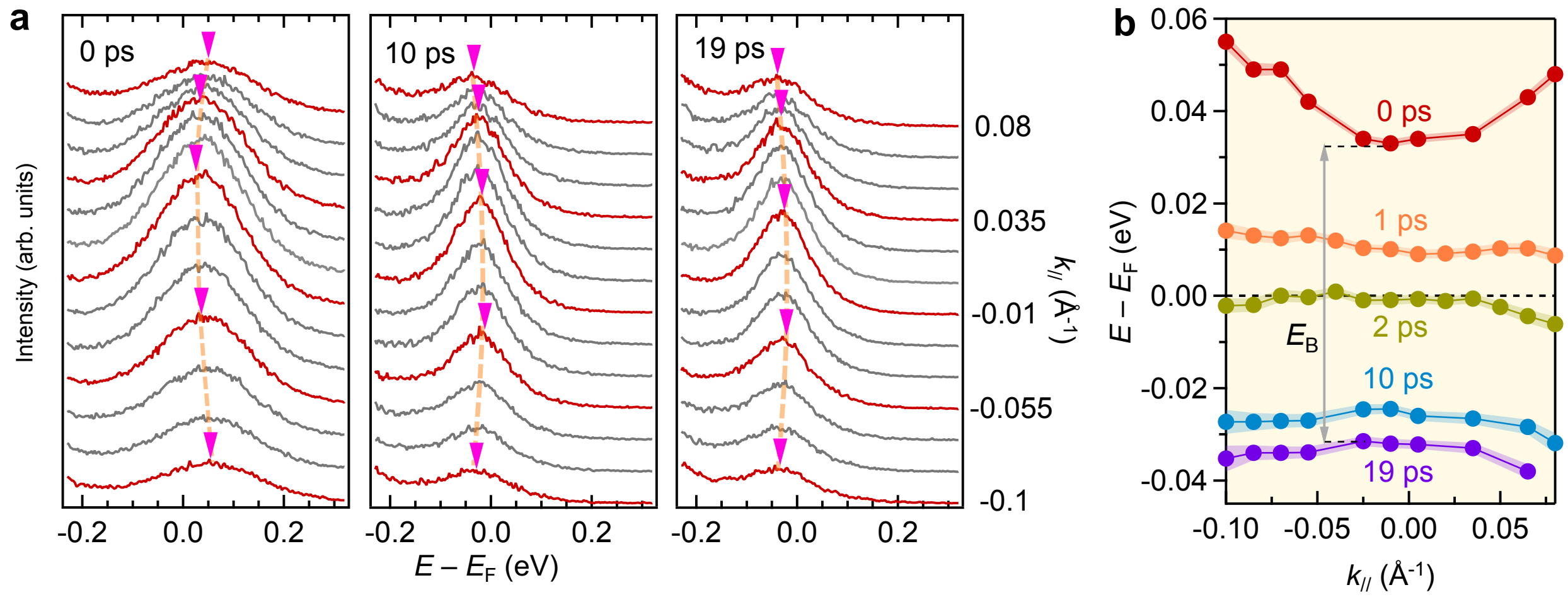

**Fig. 2| Formation of excitons evidenced by the dispersion evolution**. **a,** Energy distribution curve **(**EDC) analysis near the conduction band (CB) bottom. Tracking the EDC peaks for different momenta shows the change of the dispersion from electron-like to hole-like, representing the exciton formation. The incident pump fluence is 1.0 mJ/cm$^2$. **b,** Summary of the dispersion evolution over time. The exciton binding energy is taken as the difference between the zero-momentum feature energies at 0 ps and 19 ps. Shades represent 1σ fitting uncertainties.

We analyze the evolution of the transient band dispersion in 1-SL MBT to reveal the emergence of excitons. Figure 2**a** shows the extracted energy distribution curves (EDCs) at different momenta and representative time delays for 1-SL MBT/Si. At 0 ps, the pump photons excite electrons from the VB to the conduction band (CB) with an electron-like parabolic dispersion shown in Fig. 2**b**. This dispersion is inverted to be hole-like at later delays, evidencing the exciton formation, as shown for the 10- and 19-ps spectra. The hole-like dispersion in the trARPES spectra can be understood as the photoemitted electron mapping an exciton to the corresponding VB dispersion. It agrees with the theoretical prediction for a momentum-localized exciton distribution[40]. Notably, the hole-like dispersion of the excitonic feature reflects the VB dispersion. Previous ARPES measurements on 1-SL MBT[22,23] resolved the VB tops only near ±0.2 $Å^{-1}$, suggesting that our observed exciton feature mainly represents indirect excitons, but a finite contribution of direct excitons is possible. This excitonic feature is separated from the CB bottom observed at 0 ps by the exciton binding energy ($E_B$). For 1-SL MBT/Si, the estimation of $E_B$ is based on the spectra at 0 and 19 ps, yielding approximately 60 meV (Fig. 2**b**). Excitons in 1-SL MBT/STO, with a smaller $E_B \sim 50$ meV (Supp. Note 1), are less stable than those in MBT/Si.

The observation of exciton formation in 1-SL MBT is corroborated by distinct particle distribution statistics revealed by trARPES. In contrast to the fermion distribution in 2-SL MBT, which shows expected time-dependent cooling of a heated Fermi-Dirac distribution, the spectral distribution of 1-SL MBT shows an apparent increase of the extracted effective temperature at later delays (Extended Data Fig. 2). This counter-intuitive result can be understood through exciton broadening in the momentum space[40], and indirectly supports the assignment of exciton formation.

Intriguingly, the exciton formation in 1-SL MBT/Si is accompanied by a pronounced VB sharpening. This sharpening is revealed by differential EDCs obtained by subtracting the EDC at –2 ps from that at 2 ps (Fig. 3**b**). Following photoexcitation, a peak emerges in the differential EDC within the VB region, approximately 0.6-0.7 eV below $E_F$. Importantly, simply shifting the VB edge in the raw EDC in Fig. 3**b** does not produce the peak in the differential EDC. This peak indicates a redistribution of spectral weight beyond the physics of a dilute exciton gas. We stress that such pump-induced VB sharpening is opposite to the typical pump-induced VB broadening in other materials, particularly in systems where excitons play a dominant role in reshaping the VB structures[12,27,28]. This VB peak remains robust for at least 10 ps after photoexcitation (Fig. 4**c** and Extended Data Fig. 3), while the excitonic feature decays considerably and the transient spectral-weight depletion between 0.2-0.6 eV below $E_F$ recovers substantially. The persistence of this VB peak points to an origin different from the overall exciton population.

We investigate the origin of the transient VB sharpening through its excitation-density dependence. Using the integrated exciton spectral intensity near $E_F$ as a proxy for exciton density ($n_X$)[12], we find a quasi-linear fluence dependence for $n_X$ at low fluences (Fig. 3**d)**. In contrast, the VB peak at 2 ps is negligible below a threshold fluence $F_{th} = 0.84$ mJ.cm$^{-2}$ and emerges abruptly above it. The momentum dependence of the VB peak provides further insight. As shown in the differential ARPES spectra (Fig. 4**a**-**b**), the spectral weight loss and gain are strongest away from Γ toward the VB tops, indicating that band renormalization is the strongest near the VB maxima. Interestingly, the exciton-feature dispersion exhibits a concomitant change with fluence. At 2 ps, the hole-like dispersion for pump fluences below $F_{th}$ evolves into a camel-back shape for fluences above $F_{th}$ (Fig. 5**a**). We extract the energy shift Δ at zero momentum relative to the hole-like dispersion (Fig. S6). Δ exhibits an onset at $F_{th}$ (Fig. 5**b**), although this onset is more gradual than that of the VB peak (Fig. 3**d**). This dispersion change is also time dependent. At 4.7 $F_{th}$ and 10 ps, the exciton-feature dispersion reverts to a flatter shape with minimal zero-momentum dip (Fig. 5**a**).

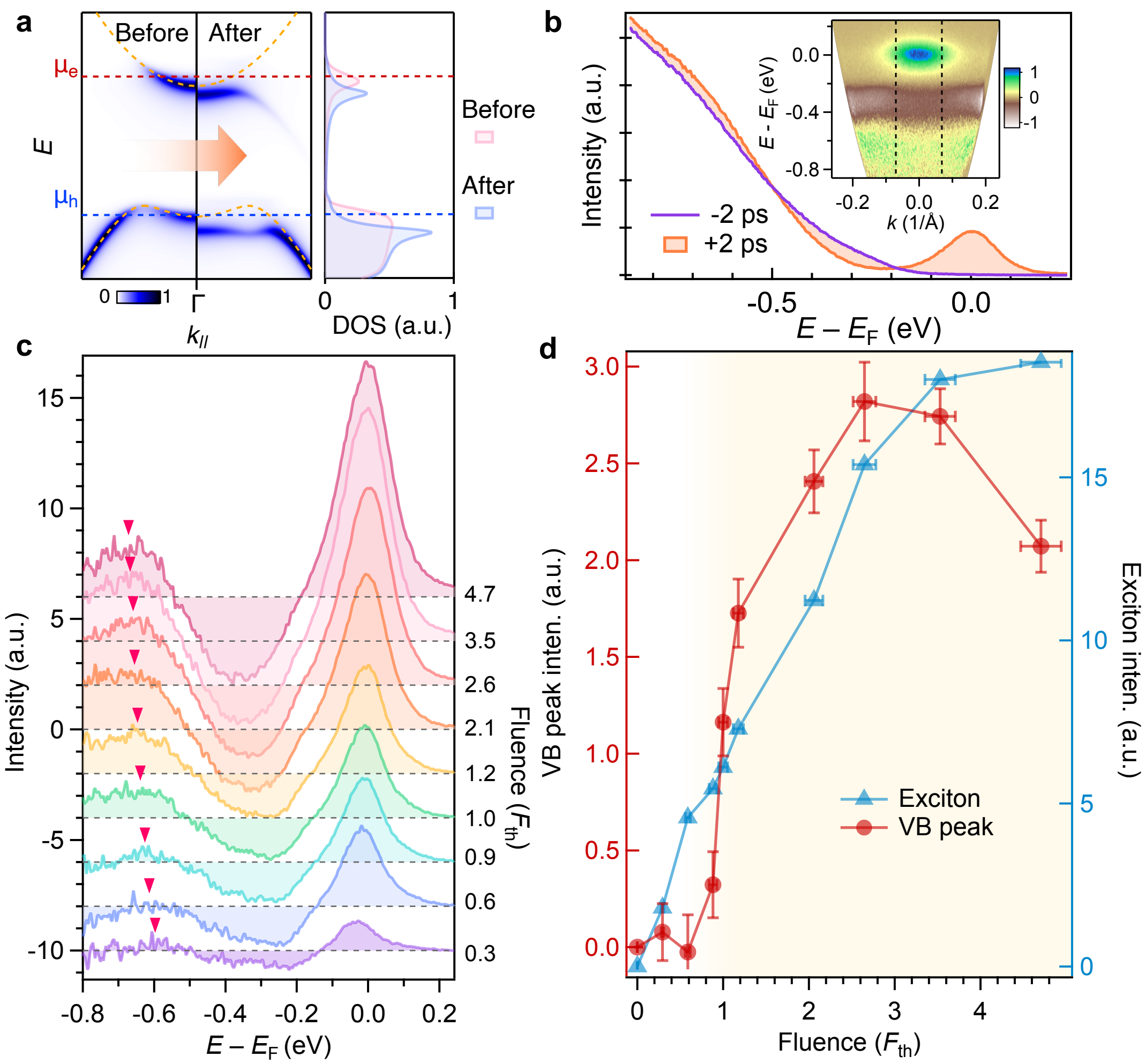

**Fig. 3| VB peak as a spectroscopic signature of exciton condensation**. **a,** A toy model for the collective exciton condensation in an MBT-like band structure. At $t_0$, free electrons and holes are excited, leading to different non-equilibrium chemical potentials. At later delays, when excitons form and condense, gaps open at these chemical potentials, leading to VB sharpening in the momentum-integrated density of states (Methods). **b,** VB sharpening revealed by comparing EDCs integrated from [-0.08, 0.08] $Å^{-1}$ around Γ before and after pumping. The pump fluence is 2.2 mJ/cm$^2$. **c,** Pump-fluence-dependent VB peak. The differential EDCs, obtained by subtracting the raw EDCs at -2 ps from those at 2 ps, are shown for various pump fluences. A peak is developed (red triangle) in the VB around 0.6-0.7 eV below $E_F$. The VB peak intensity, integrated within ±0.05 eV of the peak positions (red triangles), and the excitonic feature intensity, integrated within the intensity gain regions (> –0.2 eV), are summarized in Panel **d**. Across the threshold fluence $F_{th}$ = 0.84 mJ/cm$^2$, the VB peak exhibits an abrupt onset. The intensities corresponding to the exciton feature show an approximately linear growth across $F_{th}$ before it gets saturated at high fluences (> $4F_{th}$). Error bars show measurement uncertainties (Fig. S5). Similar results for the VB peak near -0.15 $Å^{-1}$ are shown in Extended Data Fig. 5.

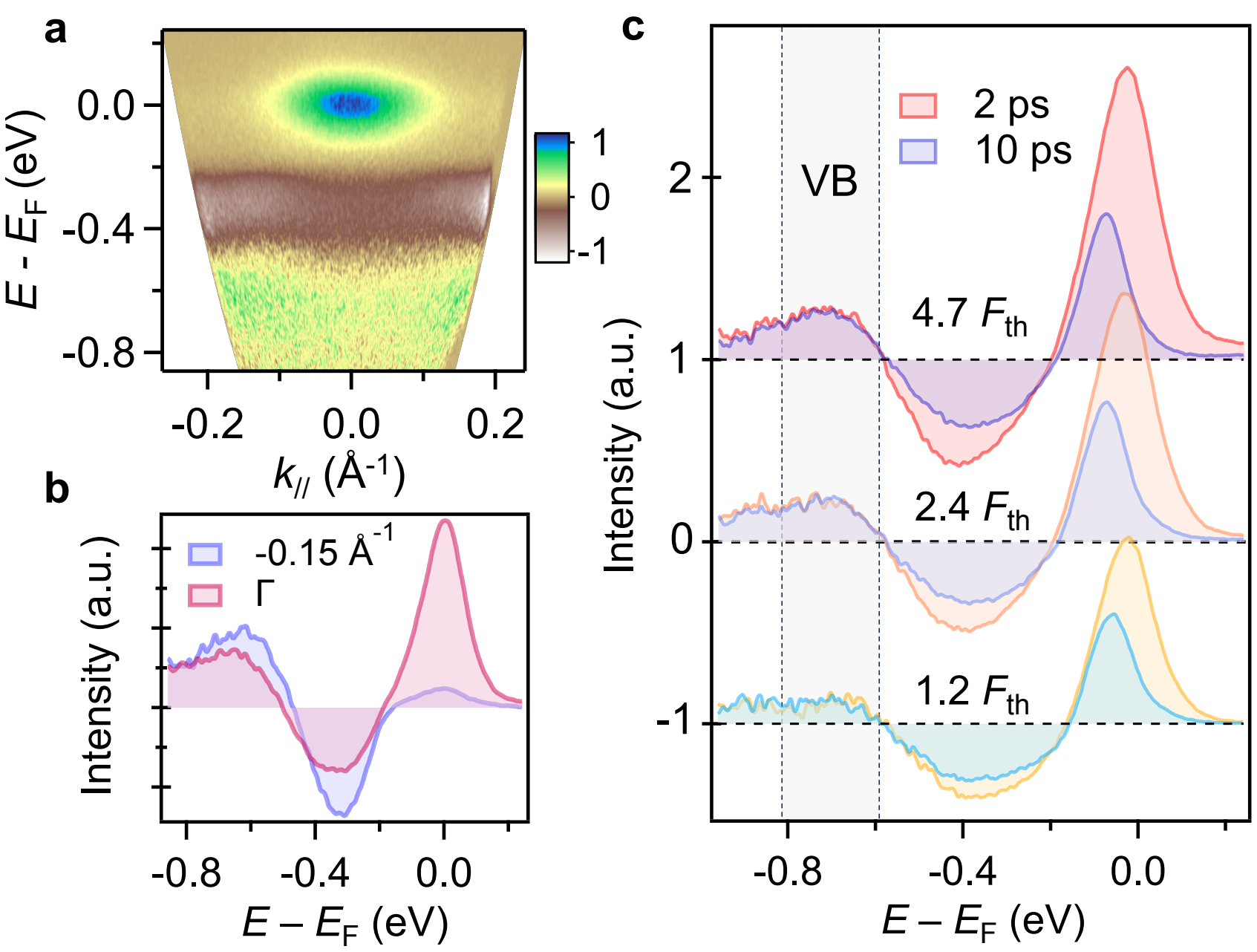

**Fig. 4| Momentum- and time-dependence of the transient VB peak. a,** The differential ARPES spectrum showing momentum-dependent VB sharpening. The spectrum is obtained by subtracting the ARPES data at -2 ps from that at 2 ps. The pump fluence is 2.2 mJ/cm$^2$. The gain and loss are strongest away from Γ, indicating that the band renormalization effect is the strongest near the VB maxima. **b**, Integrated EDCs within ±0.05 Å$^{-1}$ around Γ and -0.15 Å$^{-1}$. **c**, A distinct time scale of the VB peak dynamics compared to that of the exciton population. Even when the exciton population is substantially depleted after 10 ps, the VB peak shows minimal changes.

We summarize our main observations as follows: (i) At high $n_X$ and 2 ps, the exciton-feature dispersion evolves into the camel-back shape; band renormalization occurs near the VB tops. (ii) Both changes appear at $F_{th}$ = 0.84 mJ/cm$^2$. (iii) By 10 ps, the camel-back exciton feature relaxes, yet the VB peak is robust.

These crucial observations allow us to examine common mechanisms for transient band dispersion changes. SPV would lead to energy shifting at negative delays up to ~100 ps,[24] which is negligible in our data (Fig. S7). Optically driven Floquet[12] or Stark shifting[25] would occur only within the pump pulse (FWHM~130 fs), which cannot explain the VB peak lasting until 10 ps. The temporal onset of the VB peak closely tracks the formation of excitons on the scale of 1-2 ps (Fig. S8), which is generally faster than lattice dynamics[41]. Carrier- or exciton-screening-driven bandgap renormalizations can change the bandgap, but typically broaden the VB continuously when varying pump fluences[27,28]. Transient spin orders are unlikely, as we observe the VB peak at temperatures much higher than the Néel temperature ~15 K[42] (Fig. S9).

We turn to the most promising mechanisms: exciton-Floquet physics[12,26] and exciton condensation[15,35]. In the exciton-Floquet picture, band hybridizations between Floquet sidebands and original bands can yield camel-back exciton features and VB-top renormalization. Such hybridizations occur even for incoherent excitons[43]. In a 2D exciton condensate picture[15,44], a macroscopic coherent exciton population leads to a BKT condensate. The exciton feature's dispersion change stems from a BEC-to-BCS crossover across the condensed phase[35,36] (Supp. Note 7). The energy gap of the exciton condensate is the strongest near the VB maxima, explaining the momentum-dependent VB renormalization. Therefore, both the exciton-Floquet physics and exciton condensate can explain Observation (i).

What distinguishes these two mechanisms are the fluence- and time-dependences of our observed features. The exciton-Floquet picture is a single-particle framework. The fluence dependence of the exciton energy dip Δ and of the VB renormalization is analytical and quasi-linear[12,43] (Fig. S10). At higher fluences, the VB peak intensity is weakened[26]. These fluence-dependent behaviors do not explain our Observation (ii). Meanwhile, in exciton condensates, as the gap opens at the hole chemical potential, a strong peak emerges in the VB near the gap edge. This phenomenon was theoretically predicted for transient exciton condensation in Dirac semimetals[15], and can be compared with coherent peaks observed in superconductors due to their mathematical analogy[10]. We employ a toy model based on this theory[15] and demonstrate that a peak appears in the momentum-integrated density of states in the VB as excitons condense (Fig. 3**a**, Supp. Note 7). In a phase-fluctuating condensate framework (Supp. Note 8), the ARPES peak width in the VB is linked to the inverse phase correlation length. The BKT transition marks an abrupt change of the phase correlation function from an exponential decay to an algebraic function, resulting in the VB peak sharpening (Methods). In our experiment, such a change occurs abruptly at the NK critical density (Supp. Note 6), thus explaining Observation (ii). This sharpening in the single-particle spectral function does not apply to the exciton feature, as it represents a 2-particle gapped state (Supp. Note 8). Meanwhile, the onset of the central energy dip Δ in the exciton-feature dispersion suggests that the condensate starts in the BEC-BCS crossover regime, for which we will cross-check the validity later. This crossover regime also explains why the onset of Δ is more gradual than that of the VB peak.

Next, we examine the time dependence of our observations. In the exciton-Floquet picture, the exciton-feature dispersion and the VB renormalization both result from the Floquet-type band hybridization and should evolve together in time, which is inconsistent with Observation (iii). On the contrary, it has been demonstrated that the long-range correlation of an atomic BKT system is robust even after the boson density is quenched[45]. Thus, the BKT phase coherence, topologically protected by vortex-antivortex pairing, can outlive the depopulation mechanisms[45] such as phonon scattering and exciton recombination. This provides a natural explanation for the longer lifetime of the condensate spectral feature–the VB peak. Meanwhile, the exciton-feature dispersion (Fig. 5) reflects the BCS, BEC, or incoherent excitons tunable by the transient $n_X$, which decays according to the exciton population dynamics. For 4.7 $F_{th}$ and 10 ps, $n_X$ drops to 30% of its value at 2 ps (Fig. 4**c,** Extended Data Fig. 3). This approximately corresponds to $n_X$ obtained using $F_{th}$ and 2 ps (Fig. 3**d**), and thus explains the vanishing Δ (Fig. 5**b**).

Our observed spectral features, together with their fluence and time dependences, strongly suggest the spontaneous emergence[46] of an exciton condensate[47] (Extended Data Table 1). We provide two validity checks using parameters extracted purely from experiments. First, if the dispersion change of the exciton feature is due to the BEC-BCS crossover, then at the crossover the exciton coherence length ξ should be similar to the average inter-exciton distance $l$. Here we estimate $\xi \sim 8$ Å from the momentum spread of the exciton feature[40] (Fig. S11). At the critical fluence ~ 1 mJ/cm$^2$, we estimate $l \sim$ 3-13 Å based on near-resonant pumping on semiconductors with similar bandgaps (Supp. Note 5). Hence, near the critical fluence the system indeed enters the BEC-BCS crossover regime. Second, the superfluid density predicted by the NK formalism should correspond to the threshold $n_X$. We estimate the superfluid density based on the NK formula, the onset temperature of the VB peak (Fig. S9), and the effective mass of excitons, yielding 1.4-2.2×10$^{14}$ cm$^{-2}$ (Supp. Note 6). We also estimate the threshold $n_X$ to be ~1.6×10$^{14}$ cm$^{-2}$, based on the BEC-BCS crossover scheme ($\xi \approx l$) which occurs near $F_{th}$ (Supp. Note 6). This quantitative agreement supports our exciton condensate assignment.

With the exciton condensate picture, we sketch the phase diagram of our excitonic system in Fig. 5**b**. First, with the rapid onset of Δ above $F_{th}$, we attribute the high-density limit (>> $F_{th}$) to BCS, and the intermediate range (1.0-1.2$F_{th}$) to the crossover, as expected theoretically[35]. For the low-density limit (0.6-1.0$F_{th}$), the VB peak is detectable near the VB tops (Extended Data Fig. 4), but the exciton feature shows a negligible central energy dip (Fig. 5**a**). Notably this fluence range is below $F_{th}$, and hence macroscopic coherence is not established. Therefore, we attribute this regime to a "pseudogap BEC" (BEC*) phase, where a finite excitonic gap leads to the VB peak formation, yet long-range coherence is not achieved. Our threshold $n_X$ is ~2×10$^{14}$ cm$^{-2}$, which is 10$^2$-10$^3$ higher than that of the previous exciton-Floquet experiment on monolayer $WS_2$.[12] This marks a significant difference between the experimental schemes: the previous experiment was in the dilute, single-particle regime, while our experiment is in the dense, many-body regime.

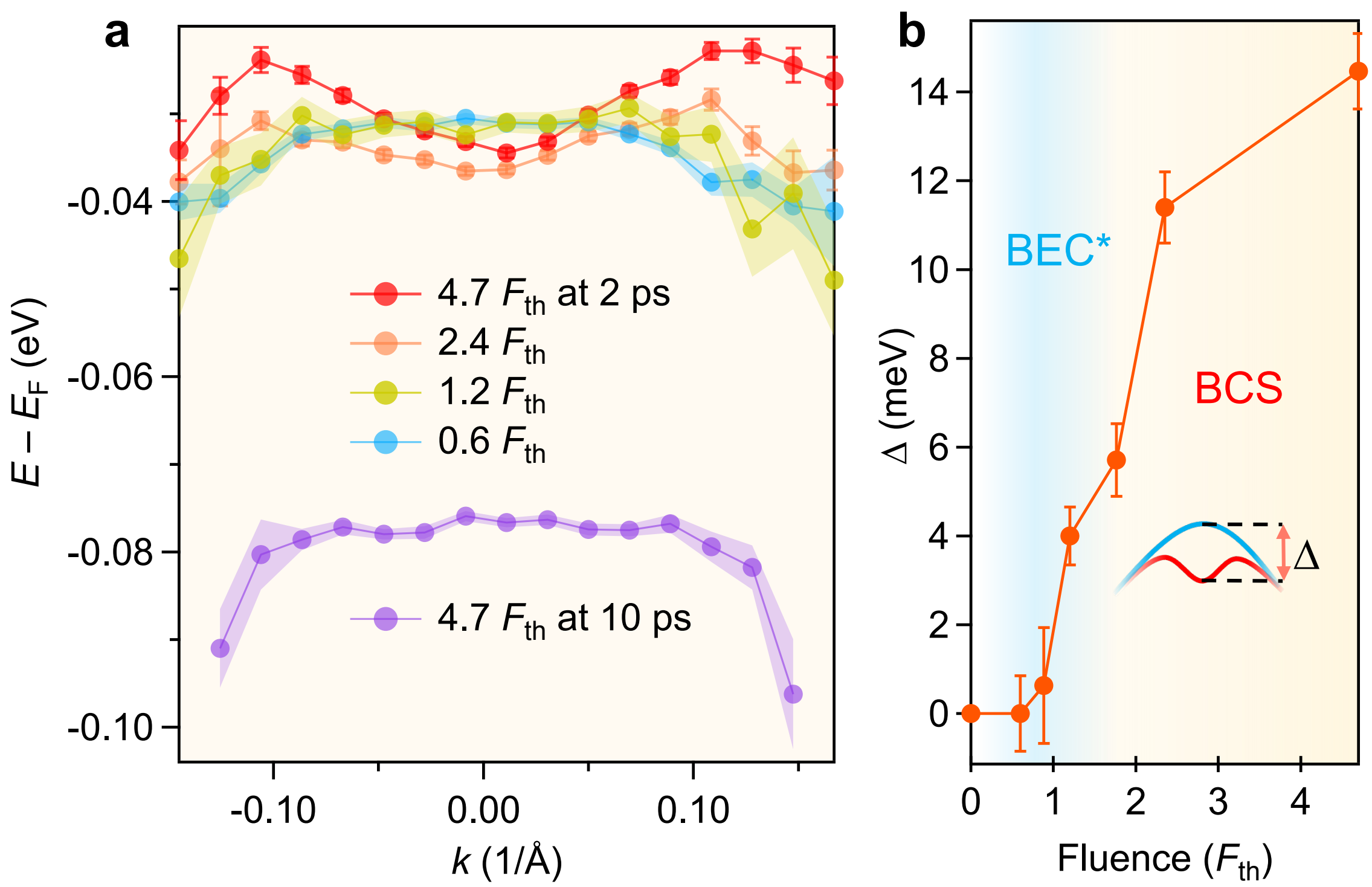

**Fig. 5| Signatures of transient BEC-BCS crossover in the exciton condensate. a,** Extracted dispersions of the excitonic feature for different excitation densities (pump fluences). Base temperature is 10 K. Extracted dispersions correspond to the BCS regime (high fluences, >> $F_{th}$), crossover regime (intermediate fluences, ~ $F_{th}$), and pseudogap BEC (BEC*) regime (low fluences, < $F_{th}$). **b,** Zero-momentum energy drop Δ versus pump fluences, while accounting for the energy shift of the excitonic feature at high momenta (≥ 0.1 and ≤ –0.1 Å$^{-1}$), as illustrated in the inset cartoon. The onset of Δ around $F_{th}$ occurs in concomitant with the VB peak onset. Error bars and shades represent the 1σ fitting uncertainties.

**Discussion**

Our work establishes a solid-state platform for observing and controlling a nonequilibrium exciton condensate and its BEC-BCS crossover. Compared with atomic boson condensates in the nano-kelvin range[48], and exciton condensates in constructed 2D materials in the milli-kelvin to kelvin range[2,3], our system provides a nonequilibrium, high-temperature analogue. Because of its analogy to superconducting condensates, this nonequilibrium exciton system may serve as a simulator of 2D superconductors. For instance, the phenomenology in the BEC-BCS crossover regime in exciton condensates may apply to 2D superconductors in the similar regime, such as monolayer FeSe/$SrTiO_3$ ($k_F\xi$ = 5-6)[49] and magic-angle twisted bilayer graphene ($k_F\xi$ = 1-10)[50], where $k_F$ is the Fermi momentum, relevant for predicting the quasiparticle band structure, pseudogap phase, and doping dependence.

Our work tests fundamental predictions of exciton condensates[14,51]. Particularly, our exciton condensate is formed in the BEC-BCS crossover regime, where the exciton coherence length is comparable to the inter-exciton distance. In the single-particle picture, the high exciton density would initialize a Mott transition that dissociates excitons into free carriers[51]. However, in the condensate framework the system evolves continuously from a condensate of bosons (BEC) to one of electrons and holes (BCS), with the order parameter remaining at similar levels[14,16,51]. When the density is further increased, the order parameter decreases[51], explaining the reduced VB peak when the fluence exceeds $4F_{\mathrm{th}}$.

Beyond its relevance to excitonic physics, our work suggests a versatile approach for engineering quantum phases inaccessible under equilibrium conditions. The coexistence of strong spin-orbit coupling, intrinsic magnetism, and optically generated excitonic order in monolayer MBT opens opportunities to explore the interplay between excitonic coherence, topology, and magnetism. Extending this approach to MBT-derived heterostructures hosting QAH, quantum spin Hall, or axion-insulating states may enable transient topological phases driven by an excitonic order. For instance, coupling two 1-SL MBT layers via an insulating buffer and applying electrical biasing may enable excitonically driven topological phases (Extended Data Fig. 6)[52]. More broadly, our work introduces a route toward dynamically manipulating correlated quantum matter for ultrafast optoelectronic and quantum applications.

## Methods

### MBE growth

Thin-film materials MBT/$SrTiO_3$ and MBT/Si were grown by molecular beam epitaxy using 99.9998% Mn, 99.999% Bi, and 99.9999% Te. 0.05 wt% Nb-doped $SrTiO_3$ (111) substrates from Shinkosha Co. Ltd. were annealed in ultrahigh vacuum (UHV) for 30 minutes at 1000 °C. Sb-doped Si(111) substrates from El-Cat Inc. was flashed at 1300 °C for 10 s and then quickly cooled down to 500 °C. The flashing was done for at least 7 times to obtain the cleanest surface for MBE growth. $Bi_2Te_3$ films were first grown on these substrates with a Bi:Te flux ratio of 1:16. The growth rate was 0.2 quintuple layer/min. The $MnBi_2Te_4$ septuple layer was formed by depositing MnTe with a Mn:Te flux of 1:20 for 7 minutes on top of $Bi_2Te_3$ and annealed at 270-300 °C in a Te-rich atmosphere. The procedure was repeated for thicker films. The films were then transferred *in situ* to the ARPES chamber for measurements.

### Static ARPES and trARPES

The static and time-resolved (tr)-ARPES measurements were performed on the Multi-Resolution Photoemission Spectroscopy (MRPES) platform at the University of Chicago[53]. The static ARPES setup features both a Helium lamp tuned to the He $1\alpha$ line at 21.22 eV and a 6 eV laser with a repetition rate of 80 MHz. The energy resolution was better than 4 meV. The trARPES setup features a Yb:KGW laser which drives a multi-stage non-collinear optical parametric amplifier (NOPA) to produce ultrafast 0.5 eV infrared pump pulses, and a single-stage NOPA along with a fourth-harmonic generator to generate 6 eV ultraviolet probe pulses. The repetition rate was 100 kHz. The energy resolution of the trARPES setup was better than 20 meV. Focused 6 eV probe beam waists, as characterized by the full-widths-half-maxima (FWHMs), were $13\times14$ $\mu m^2$ and $20 \times 30$ $\mu m^2$ for the static and time-resolved ARPES experiments, respectively. The FWHM of the pump beam waist was $105 \times 120$ $\mu m^2$. A systematic alignment procedure was adopted to ensure the overlap of the probed regions for static and time-resolved ARPES[39]. The linearly s-polarized, ~130 fs-long IR pump pulses were focused to produce incident fluences of 3.96 mJ/$cm^2$ and below. The overall time resolution was determined to be ~200 fs.

### Transmission Electron Microscopy (TEM)

TEM Lamella were prepared in a Thermo Fisher Helios 5CX cryo-scanning electron microscope with focused ion beam (cryoFIB-SEM) dual-beam. Thermo Fisher Scientific MAPS 3.3.1 was used to stitch an overview of the lamella for site selection. MBT/$SrTiO_3$ and MBT/Si samples were cut along the $[11\bar{2}]$ direction of the substrates. Protective coating was deposited with a gas injection system in four steps to ensure minimal damage to the film. 300 ns dwell time and 50% overlap at 52 degrees tilt was used throughout the deposition. First step was electron-beam-assisted carbon deposition, followed by electron-beam-assisted tungsten deposition. Ion-beam-assisted tungsten deposition was conducted in two steps to increase the protective layer thickness. Trenches were drilled and the lamella was extracted with a 16 kV $Ga^{+}$ ion beam to a 3 mm copper half grid. The lamella was thinned at 5 kV ion beam voltage to a final thickness of 120 nm and polished with 2 kV.

Analytical Transmission and Scanning Transmission Electron Microscopy (TEM/ STEM measurements in this study were conducted using the Analytical PicoProbe Electron-Optical Beam Line / Iliad Ultra Scanning Transmission Electron Microscope at Argonne National Laboratory. Lamellas were loaded into the microscope with a side entry UltraX Thermo Fisher Scientific double-tilt holder and oritented to the desired zone axis in TEM (Fig. S13). Imaging reported herein was conducted at 300 keV with a probe current of 100 pA using a Ceta-2 CMOS detector, and the Integrated Differential Phase Contrast (iDPC) STEM using the microscope's integrated Panther detectors. HRSTEM images were 2048 x 2048 with a pixel size of 5.08 pm and a dwell time of 3.2 µs. Throughout, no significant hydrocarbon contamination was observed that would affect measurements.

## Analytical model of the exciton condensation and VB peak

We adopt the analytical model by Triola *et al.* describing excitonic condensation in 2D Dirac bands[15] and modify it for our MBT-like dispersions with a parabolic CB and an M-shaped VB, separated by an indirect bandgap $E_g$, as shown in Fig. 3**a**. As a toy model aiming to explain the basic qualitative observations, it assumes the contact-interaction limit. This is similar to the weak-coupling limit of the BCS theory, resulting in a uniform order parameter across all momenta which is restricted to a thin momentum shell near the non-equilibrium chemical potentials. A finite-$\boldsymbol{Q}$ excitonic order is enforced, based on the momentum separation between the CB bottom and the VB tops.

The most important quantity in the simulation of the spectral function is the self-energy $\Sigma_{v(c)}(\boldsymbol{k}, \omega)$ for the VB (CB) electrons at momentum $\boldsymbol{k}$ and energy ω, which is obtained as

$$\Sigma_{v(c)}(\boldsymbol{k}, \omega) = \frac{|\delta_{\boldsymbol{k}}|^2}{\omega - \xi_{c(v)\boldsymbol{k}\pm\boldsymbol{Q}}} \approx \frac{|\delta_{\boldsymbol{0}}|^2}{\omega - \xi_{c(v)\boldsymbol{k}\pm\boldsymbol{Q}}}, \quad (1)$$

with order parameter $\delta_{\boldsymbol{k}} \approx \delta_0$ and band dispersions $\xi_{c(v)\boldsymbol{k}}$. The self-energy renormalizes the VB dispersion causing a spectral weight redistribution after the excitons condense, via the retarded Green's function of the VB, $G_v^R(\boldsymbol{k}, \omega)$, as:

$$G_v^R(\boldsymbol{k},\omega) = \frac{1}{\omega - \xi_{v\boldsymbol{k}} + i0^+ - \Sigma_v(\boldsymbol{k},\omega)}. \tag{2}$$

This model phenomenologically assumes a uniform order parameter $\delta_0 = E_g/10$. The occupied density of states is imposed by a FD distribution at each electron or hole chemical potential at $T = 100$ K. More details of the derivation and deployment of this model are presented in Supp. Note 7.

Beyond the BCS-like mean-field treatment, a statistically phase-fluctuated excitonic order is also modelled based on a phase-fluctuation theory of 2D superconductivity[29], but modified for the case of exciton condensates. This is feasible due to the analogous mathematical formalism of the electron self-energy in both models. A finite phase-correlation-length energy scale $\gamma_\phi(\boldsymbol{k})$ enters the self-energy of the VB in Eq. (1) as:

$$\Sigma_v(\boldsymbol{k},\omega) \cong \frac{\delta_0^2}{\omega - \xi_{c\boldsymbol{k}\pm\boldsymbol{Q}} + i\gamma_\phi(\boldsymbol{k}+\boldsymbol{Q})}, \tag{3}$$

which subsequently broadens the VB peak and weakens its intensity. At the macroscopically coherent limit: $\gamma_\phi(\boldsymbol{k}) \to 0$, the self-energy reduces to Eq. (1). This model predicts an increasingly sharpened VB peak as the phase correlation length grows ($\gamma_\phi(\boldsymbol{k})$ decreases), and a vanishing peak with a remnant pseudogap when the phase correlation length is small enough. Theoretically, the phase correlation length rapidly diverges as $\exp\left(\frac{b}{\sqrt{T_{BKT}/T-1}}\right)$ when the system enters the BKT phase, transforming the phase correlation function into an algebraic form $\langle \exp[i\theta(\boldsymbol{r}) - i\theta(0)] \rangle \sim \left(\frac{a}{r}\right)^{\frac{mk_BT}{2\pi n_s \hbar^2}}$.[54] Here $T_{BKT}$, $T$, $\theta$, $\boldsymbol{r}$, $m$, $n_s$ stand for the BKT temperature, system temperature, phase of the order parameter, two-point distance, exciton mass, and superfluid density; $a$ and $b$ are constants. Following this framework, our observed sudden onset of the VB peak sharpening at the threshold fluence can be related to the appearance of a long-range coherent exciton condensate, reflecting the NK jump. Furthermore, the distinctively long relaxation time of the VB peak suggests that the phase coherence stays longer than the overall exciton population, in agreement with observations in atomic BEC systems[45]. More details of the derivation and application of this model are presented in Supp. Note 8 and Fig. S12.

**Author contributions**

Supervision: SLY

Conceptualization: SLY, KDN

MBE growth: KDN, YB, WL

ARPES measurements: KDN, GB, YB, WL, HL, JY

Theoretical models and calculations: KDN, SLY

Data analysis and visualization: KDN

STEM characterization: TM, YSM, CL

Writing—original draft: KDN, SLY; Writing—review & editing: All authors

**Competing interests:**

The authors declare no competing interests.

**Acknowledgements:**

The authors thank D. Qiu and J. Guo at Yale University, K. Levin, P. B. Littlewood, and J. Park at the University of Chicago, Z.-X. Shen at Stanford University for helpful discussions. The molecular beam epitaxy growth, static angle-resolved photoemission spectroscopy (ARPES) characterization, and scanning transmission electron microscopy (STEM) on $MnBi_2Te_4$ films were supported by DOE Basic Energy Sciences under Grant No. DE-SC0023317. The time-resolved ARPES study was supported by DOE Basic Energy Sciences under Grant No. DE-SC0022960. The mid-infrared generator was supported partially by the Gordon and Betty Moore Foundation, Grant No. GBMF12763.